\documentclass{Interspeech}

\interspeechcameraready

\title{Automated evaluation of children's speech fluency for low-resource languages}

\author[affiliation={1,2}]{Bowen}{Zhang}
\author[affiliation={1}]{Nur Afiqah Abdul}{Latiff}
\author[affiliation={1}]{Justin} {Kan}
\author[affiliation={1}]{Rong}{Tong}
\author[affiliation={1}]{Donny}{Soh}
\author[affiliation={1,3}]{Xiaoxiao}{Miao}
\author[affiliation={1}]{Ian}{McLoughlin}

\affiliation{ICT Cluster}{Singapore Institute of Technology}{Singapore}
\affiliation{College of Computing \& Data Science}{Nanyang Technological University}{Singapore}
\affiliation{Division of Natural and Applied Sciences}{Duke Kunshan University}{China}

\email{bowen009@e.ntu.edu.sg, 
\{nurafiqah.abdullatiff,
justin.kan,
tong.rong,
donny.soh,
xiaoxiao.miao,
ian.mcloughlin\}@singaporetech.edu.sg
}
\keywords{speech recognition, human-computer interaction, computational paralinguistics}

\usepackage{comment}
\usepackage{multirow}

\usepackage{todonotes}
\usepackage{setspace}

\begin{document}

\maketitle

\begin{abstract}
    
Assessment of children's speaking fluency in education is well researched for majority languages, but remains highly challenging for low resource languages. This paper proposes a system to automatically assess fluency by combining a fine-tuned multilingual ASR model, an objective metrics extraction stage, and a generative pre-trained transformer (GPT) network. The objective metrics include phonetic and word error rates, speech rate, and speech-pause duration ratio. These are interpreted by a GPT-based classifier guided by a small set of human-evaluated ground truth examples, to score fluency. We evaluate the proposed system on a dataset of children's speech in two low-resource languages, Tamil and Malay and compare the classification performance against Random Forest and XGBoost, as well as using ChatGPT-4o to predict fluency directly from speech input. Results demonstrate that the proposed approach achieves significantly higher accuracy than multimodal GPT or other methods.
\end{abstract}

\section{Introduction}
In a multi-ethnic, multi-lingual society such as Singapore, or in mixed-race households, children may not be exposed to their ``mother tongue'' at home. 
To preserve cultural heritage and promote bilingualism, the Singapore Ministry of Education ensures that all primary school students learn their mother tongue, namely Mandarin, Malay, or Tamil, alongside English, which is the primary language of instruction and daily interaction.
However, traditional methods of assessing oral fluency in these mother tongues often rely on human evaluation, which can be time-consuming, subjective, and inconsistent.
Significant research into areas such as Computer-Assisted Language Learning (CALL) \cite{call}, offers potential solutions by integrating technology into language teaching and learning. CALL systems provide self-paced, interactive learning experiences and automated evaluation, with immediate feedback on pronunciation, fluency, and accuracy. 

Among various assessments, researchers have developed several approaches to deriving fluency scores, which are key metrics to reflect the smoothness, naturalness, and efficiency of speech production. The methods are broadly categorised into non-ASR and ASR-based methods. The former aims to assess fluency without relying on ASR. For example, in~\cite{ASR_free_IS2020}, a generative model was trained on the marginal distribution of speech, where fluency is assessed by comparing raw speech without recognising specific words or phonemes. Similarly, self-supervised speech representation schemes, such as wav2vec~\cite{wav2vec2}, have been explored. These methods leverage pre-trained acoustic models to infer fluency-related metrics directly from speech, such as rhythm, pause patterns, and speech rate. While non-ASR models can be computationally efficient and less reliant on annotated data, they often lack the granularity to provide detailed insights into specific fluency issues, such as word-level disfluency or pronunciation errors.

In the realm of language learning, ASR-based methods have become dominant in recent years, where speech first transcribed using an ASR model can be analyzed for linguistic accuracy, fluency, and pronunciation~\cite{matsuura2022refinement, wang2018towards}. Key metrics include WER (Word Error Rate) for transcription accuracy, phoneme alignment for pronunciation, speech rate and pauses for fluency, and prosodic features (pitch, intensity, rhythm) for naturalness~\cite{bamdev2023automated}.
Though ASR-based solutions show promising performance for adult speech in high-resource languages such as English and Chinese, they face major challenges related to the variable nature of children's speaking patterns, leading to low baseline accuracy for children's speech.
The scarcity of high-quality annotated children's speech data, compounded by ethical concerns around data collection, pose additional obstacles.
The challenges are even greater for low-resource languages.
\begin{figure}[!t]
    \centering
   
    \includegraphics[width=0.9\linewidth]{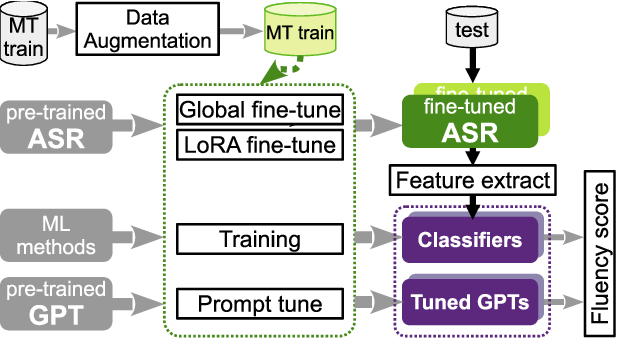}
    \caption{Proposed automatic fluency scoring framework showing adaptation of pre-trained ASR and GPT models using highly augmented low-resource mother tongue (MT) language data.
    }
    \vspace{-7mm}
    \label{system}
\end{figure}
 
In this paper, the challenges of developing low resource automated assessment systems for children's mother tongue speech fluency are addressed specifically for the two lowest resource MLs in Singapore: Malay and Tamil. 
We propose an ASR-based language proficiency assessment system as illustrated in Fig.~\ref{system} and evaluate the proposed system for Malay and Tamil, demonstrating very good performance. 
These findings are expected to be useful for speech evaluation in other low-resource language scenarios.
Specifically, to tackle training data scarcity, we adapt and fine-tune a high-quality multilingual adult ASR engine using a small set of labeled child speech data.
From the fine-tuned ASR, we extract metrics such as phonetic and word error rates, speech rate, and speech-pause duration ratios. A GPT network is then prompt-tuned to generate fluency prediction scores based on these metrics. The system is optimized to align with a limited sized human-evaluated ground truth training dataset (discussed in more detail in Section \ref{sec:child}).

The system is effective because it not only successfully harnesses the established capability of the pre-trained multilingual adult ASR engine, but also makes use of powerful generalisation and flexible adaptation properties of the GPT network that can capture the nuances of human-like assessment decisions.

\section{Related works}
\label{sec:related}
Speech fluency evaluation is inherently subjective and challenging to automate. 
This section revisits related work on dominant ASR-based evaluation models, which lead the field due to their capacity for fine-grained speech analysis.

These ASR-based methods transcribe learner speech into text, and also enable extraction of detailed metrics such as word error rate (WER), phoneme error rate (PER), and pause duration. For example, Goodness of Pronunciation (GOP)~\cite{Witt_GOP} is widely used to detect mispronunciations, which can significantly impact fluency. Recent advancements, such as Transformer models with multi-task learning from GOP features~\cite{GOPT_ICASSP2022}, have achieved promising results in fluency assessment for non-native speakers. Additionally, ASR-based systems can capture prosodic information like pitch, intensity, and rhythm, which are essential for evaluating the naturalness of speech.
With the emergence of Whisper~\cite{Whisper}, more automated DNN driven speech evaluation methods have been developed.
For example, MultiPA \cite{chen2024multipa}, a multitask pronunciation assessment model, extracts phone alignment features from Whisper and word embeddings from HuBERT \cite{hsu2021hubert} to enhance assessment accuracy. MOSA-Net+ \cite{zezario2024study} enhances MOSA-Net by incorporating Whisper acoustic features and self-supervised learning (SSL) embeddings for more accurate speech quality prediction.

While ASR-based models excel in adult speech evaluation for common languages, they face significant challenges when applied to children's speech, or to low-resource languages.
This is because both the front-end ASR model, which extracts fine-grained features for fluency evaluation, and the back-end DNN-based classifiers, which assess fluency scores based on these fine-grained features, require large-scale high quality labelled data for training.
In the case of children's speech, the variability in pitch, pronunciation, and speech patterns differs significantly from adult speech, making it difficult for models trained on adult data to generalize effectively. To address these challenges, researchers have explored various strategies. For instance, a multitask learning framework was proposed in~\cite{multitask_IS2017} to model the characteristics of both adult and child speech, improving recognition accuracy for children. A FBDS-based automatic
acoustic measure was proposed for Korean children speech
fluency prediction~\cite{fontan2022predicting}. Data augmentation techniques, such as pitch, speed, tempo, volume perturbation~\cite{dataaug_2021}, have also been employed to simulate children's speech and expand the available training data. 
Low-resource languages also lack sufficient labelled data, hindering the ability of ASR systems to learn the necessary linguistic features and patterns, leading to suboptimal performance. These challenges highlight the need for specialised datasets and tailored approaches to improve ASR accuracy and robustness in these domains.
Several adaptation approaches have been tried, for example in low-resource Chinese children ASR~\cite{S2_Lora}. Fine-tuning, by adapting the final layers of Whisper or Wav2vec2 models with children's speech data, has shown promise at improving ASR performance for low-resource languages~\cite{finetune_whisper_children}, so we also adopt this method.

\section{Children speech evaluation}
\label{sec:child}
To overcome the poor generalization of adult ASR for children's speech, we collected a dataset of children's Malay and Tamil speech for fine-tuning an adult ASR model.
We leveraged our fine-tuned model to extract various objective metrics. %
These then served as input features for comparing traditional machine learning models and the proposed GPT-based approach to assess children's speech fluency in both languages. The following sections discuss further.

\subsection{Dataset collection and annotation}
\label{sec:dataset}
To address the lack of publicly available children's speech datasets for Malay and Tami, we collected an in-house dataset. What is unusual is the inclusion of both reading and question-answer (Q\&A) tasks for each language. 
The dataset comprises speech samples from 218 male and 436 female primary one and primary two students in Singapore. In total, 654 recording sessions were conducted in real classroom environments. Data collection was approved by the Institutional Review Board of our university and by the Ministry of Education.

The reading task involved students reading 27 sentences aloud line by line. The Q\&A task involved a picture description section, where students observed 6 pictures and answered 4 to 5 related questions per picture within a 3-minute timeframe. Each session was recorded, capturing both reading and conversational responses. 
Due to the real classroom settings, the recordings contain considerable ambient background noise.

Recordings were segmented into utterances, which were then transcribed by human experts to ensure accurate annotations. Each sentence was also assigned a fluency score on a scale of 1 to 4, with 4 being the highest. Since fluency scores of 1 and 2 were underrepresented in the collected data, we merged them into a single category labeled ``low''. Fluency score 3 is labeled ``medium'', and fluency score 4 is labeled ``high''. To avoid nonsense input, we selected the medium and high fluency sentences only for ASR fine-tuning, but used all for fluency model training and classification.

\begin{table}[tb]
\caption{Number of utterances used to fine tune ASR models}
\vspace{-4mm}
\begin{center}
\label{dataset_ASR}
\scalebox{0.9}{
\begin{tabular}{clll}
\textbf{Split} & \textbf{Number of utterances} & \textbf{Malay} & \textbf{Tamil} \\ \hline
\multirow{3}{*}{\textbf{Train}} & Children            & 5853  & 2037 \\
                                & Children + \textit{AugA} & 7359  & 6031 \\
                                & Children + \textit{AugB} & 18329 & 6837 \\ \hline
\multirow{2}{*}{\textbf{Test}}  & Reading             & 178   & 114  \\
                                & Picture             & 200   & 210  \\ 
\end{tabular}
}
\end{center}
\vspace{-8mm}
\end{table}

\subsection{Data augmentation}
To address the scarcity issue, we employed two data augmentation methods. \textit{AugA} utilised publicly available adult speech data to which we applied vocal tract length normalisation (VTLN) to simulate children's vocal characteristics~\cite{m2016mismatched}. Children have shorter vocal tracts, resulting in higher formant frequencies. Bandpass filtering, formant shifting, pitch shifting (to 150Hz for male, 250Hz for female samples), and speed change (0.8) were also applied as in~\cite{dataaug_2021}. \textit{AugB} varied the speech rate, pitch and intensity for both children and adults' data to increase the volume of data for fine-tuning. 
Table \ref{dataset_ASR} summarises the utterances per language, where the ``Children'' subset contains only medium and high fluency utterances, and either \textit{AugA} or \textit{AugB} substantially increases the data quantity.
There was no speaker overlap between train and test sets.

\subsection{Low resource Children's fine-tuned ASR}
\label{sec:childrenASR}

To obtain the fluency-related metrics, we used the multilingual Whisper~\cite{Whisper} large ASR model as a baseline. It was trained on 680,000 hours of weakly supervised speech and supports over 100 languages, making it a strong foundation for our task. For better performance on Malay and Tamil, we also utilized the open sourced pre-finetuned Whisper models for these two languages respectively.

To adapt to children's speech, we fine-tuned the model parameters using the target children's speech dataset described in section \ref{sec:dataset}. Given Whisper’s strong latent language representations, fine-tuning enables the model to better capture pronunciation variations and fluency patterns in the target speech.
To reduce training complexity, we employed Low-Rank Adaptation (LoRA)~\cite{LORA}. %
LoRA introduces trainable adapter layers while keeping the original model weights frozen, significantly reducing the number of trainable parameters. Instead of updating the entire weight matrix, LoRA factorizes it into smaller matrices, allowing efficient adaptation to children's speech.

\subsection{Fluency scoring metrics}
\label{sec:features}

Fluency scoring evaluates how smoothly and naturally a speaker produces language and typically is highly subjective. 
Aiming to automate the assessment, we extracted metrics from the fine-tuned ASR model output as summarised in Table~\ref{features} 

\begin{table}[htbp]
\caption{Metrics for automatic fluency assessment}
\vspace{-4mm}
\begin{center}
\scalebox{0.9}{
\begin{tabular}{ll}
\multicolumn{2}{l}{\textbf{Primary metrics}}\\
\hline
 Language & Malay or Tamil \\
 Task type  & Reading (R) or Picture Q\&A (P) \\
 WER   & word error rate               \\
 CER   & character error rate          \\
 PER   & phoneme error rate               \\
 Pause duration &  duration between words\\
 Total duration & total duration of speech                    \\
 Num pauses     & number of pauses in the sentence  \\
 \multicolumn{2}{l}{\textbf{Secondary metrics}}\\
\hline
 Speed  & \begin{tabular}[c]{@{}l@{}} num\_words/total\_duration$\times$60\end{tabular} \\
 Pause ratio    & pause\_duration/total\_duration       \\ 
\end{tabular}
}
\vspace{-3mm}
\label{features}
\end{center}
\end{table}
This included four granularities for accuracy at utterance, word, character, and phoneme level. 
Errors in a single syllable can significantly impact both the character error rate (CER) and word error rate (WER), especially in phonetically based languages like Tamil, where small syllable changes can radically alter word meanings. The phoneme error rate (PER) is computed by converting words into their International Phonetic Alphabet (IPA) equivalent~\cite{CUPbook2016}, to provide a more precise evaluation of pronunciation accuracy.

Speech rate is another key indicator of fluency. A consistently slow rate may suggest difficulty with language retrieval or articulation, while an overly fast rate can indicate disfluency due to unclear articulation or skipped words.

Speech pauses provide insight into cognitive processing and articulatory control. While natural pauses aid comprehension, excessive or poorly timed pauses may indicate fluency challenges. A high pause ratio may suggest language difficulties or lack of confidence. To capture these patterns, we measure the proportion of time spent pausing relative to total speaking time, along with speech rate and error rates at different linguistic levels (word, character, phoneme).

\subsection{Fluency score prediction model}

Due to the scarcity of children's data, especially in low-resource languages, training a deep neural network (DNN) adequately for speech evaluation is challenging. 
However, given our set of metrics and small but well labelled dataset, %
we can leverage traditional machine learning methods, such as random forest and XGBoost.
We also compared performance against a multi-modal LLM, which can input both raw speech and text prompts. In this case we used prompt tuning to obtain fluency scores from metrics plus the raw speech.

\begin{figure}[tbp]
    \centering
    \includegraphics[width=0.9\linewidth]{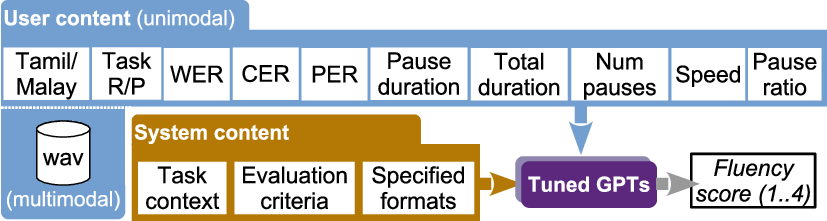}
    \caption{Automatic scoring using tuned GPTs. The fixed ``system content'' defines the task, the context and the input/output formats. ``user content'' contains per-instance input data.} %
    \label{workflow}
    \vspace{-5mm}
\end{figure}

\subsubsection{Traditional machine learning methods}
We evaluated two machine learning methods for automatic fluency scoring: Random Forest and XGBoost. They both rely on ensemble learning techniques to improve prediction accuracy.

\subsubsection{GPT-based models}
We also compare two GPT-assisted approaches.
The first is a multimodal LLM, gpt-4o-audio-preview\footnote{https://platform.openai.com/docs/models\#gpt-4o-audio}, with direct speech inputs and a prompt-tuned fluency scoring task.

The second is gpt-4o-mini\footnote{https://platform.openai.com/docs/models\#gpt-4o-mini} with text-only input, prompt-tuned as a meta-evaluator based on our extracted metrics, context, and prototypes. 
Both models were pretrained on vast amounts of language data, which allows them to understand patterns in pronunciation, rhythm, and speech rate. The audio-input model is additionally trained on a large selection of speech samples, including speech in both languages. 

When using LLMs, prompt design is crucial, and this is shown in Fig.~\ref{workflow}. 
The input prompt is composed of two main components, the ``system'' and ``user content''. 
The first outlines the task context, specifies the evaluation metrics, defines the input-output format, and most importantly, provides prototypes and boundaries for the evaluation process. 
The second, ``user content'' is structured as a JSON object that encapsulates all necessary features for the speech evaluation.
For the unimodal system, this included the metrics extracted from our fine-tuned Whisper models for each language, and contained speech utterance files for the multimodal system. 
The output is a quantitative fluency score for each utterance.

\section{Experiment} 
\label{sec:setup}

We conducted extensive experiments to obtain an optimal ASR model. Whisper-medium for Malay and Whisper-small for Tamil were adopted for the full fine tuning, sized to match the data availability (much more training data for Malay than Tamil).
We also leveraged two pre-finetuned Whisper models: Mesolitica-medium\footnote{https://huggingface.co/mesolitica/malaysian-whisper-medium} for Malay and Vasista-medium\footnote{https://huggingface.co/vasista22/whisper-tamil-medium} for Tamil as base models prior to the LoRA fine tuning.
For fine-tuning, we used the datasets described in section \ref{sec:child}. 

The best ASR model in terms of picture task WER (prioritised for its broader vocabulary and realistic context) was adopted to extract the metrics described in Section \ref{sec:features}. %
The machine learning speech evaluation models\footnote{https://scikit-learn.org/1.5/modules/generated/sklearn.ensemble. RandomForestClassifier.html} \footnote{https://xgboost.readthedocs.io/en/stable/parameter.html} were trained on metrics extracted from all utterances using default parameters. 

\subsection{ASR model fine-tuning performance}
\label{sec:exp}

\begin{table}[tb] %
\caption{ASR performance on various models} %
\vspace{-5mm}
\begin{center}
\tabcolsep=0.15cm
\scalebox{0.9}{
\begin{tabular}{llcccc}
\multicolumn{2}{l}{\textbf{Language/WER(\%)}}&\multicolumn{2}{c}{\textbf{Malay task}} & \multicolumn{2}{c}{\textbf{Tamil task}} \\ \hline 
Model & Training & R & P & R & P \\ \hline
Whisper-sm & NA &71.16&	72.28	&86.33	&86.67\\ 
Whisper-med & NA &60.96	&66.30&	80.09	&78.94\\
Whisper-lg\_{v3} & NA &55.81	&59.21	&79.12	&76.96 \\ 
Pre-finetuned & NA & 53.62 & 46.14 & 40.48 & 50.43 \\ \hline
Global & {Children} & 7.46	&13.86	&40.43	&75.07 \\ %
fine-tune& {~~~+AugA} & 7.68	&\textbf{13.70}	&29.12	&52.02  \\ %
& {~~~+AugB} & 8.00	&\textbf{13.70}	&\textbf{24.41}	&48.46  \\ \hline
LoRA  & {Children} &9.54	&15.28	&39.11	&52.19 \\ %
fine-tune& {~~~+AugA} & 10.09	&16.06	&30.75	&45.59  \\ %
& {~~~+AugB} & \textbf{7.24}	&14.31	&32.90	&\textbf{40.06} \\ 
\end{tabular}
}
\vspace{-2mm}
\label{fine-tuning}
\end{center}
\end{table}

First we assess the ASR performance we were able to achieve for both Malay and Tamil in Table \ref{fine-tuning}. A lower WER indicates better model performance. R is the reading task and P is the picture Q\&A task.
The performance of the fine-tuned models is shown using different different datasets and augmentations (Aug). 
All base models performed poorly on children's speech, with Tamil performing worst. 
This is unsurprising because Tamil ASR has long been challenging due to limited training data, and children's Tamil ASR even more so. %
Results show that both global fine tuning and LoRA fine tuning significantly improve performance, with Malay enjoying the greatest gain, likely due to having more fine tuning data. 

For Tamil, global fine tuning was more effective on the R task, while the LoRA-finetuned model performed better on the P task. 
This may be because reading data has a more limited vocabulary, causing full fine tuning to overfit and become less generalised. %
Overall, data augmentation improved model performance, with one outlier being the LoRA-fine tuned model for Tamil task R. %
By analysing specific recognition results, we found that the LoRA model fine tuned on Children+AugB sometimes incorrectly combined or separated words. %

\subsection{Fluency assessment}

Next we assess our automated scoring framework via Pearson correlation, balanced accuracy~\cite{balancedAcc, balanced2}, and weighted F1 score (needed due to the imbalanced, multi-class nature, of the training data).  
Table \ref{fluency_result} presents fluency prediction results. 
Among machine learning methods, XGB achieved the best performance for both Malay and Tamil. 

However, both GPT systems outperformed traditional methods for both Malay and Tamil, which is impressive given that the GPTs are essentially performing zero-shot inference unlike machine learning models that required training. 
The superior performance for Malay over Tamil can be directly attributed to its more effective ASR model, which yields more accurate acoustic and tempo metrics than the Tamil model. %
The multimodal LLM, gpt-4o-audio-preview{$^\dagger$}, which directly inputs speech waveforms, performed less well indicating that current LLM models may not be well trained for Tamil or Malay speech. This is likely to be true as long as both remain low-resource languages.
Our proposed system (gpt-meta), outperformed all other methods by a significant margin. For Malay speech it achieved a highly impressive F1 score of 0.91, and it also outperformed all other systems for Tamil speech.

\begin{table}[tb]
\caption{Results for RF and XGB; Corr: Pearson correlation; Acc: accuracy (\%); F1:weighted F1 (\%).}
 \vspace{-3mm}
 \label{fluency_result}
 \begin{center}
\scalebox{0.9}{
\begin{tabular}{lcccccc}
{} & \multicolumn{3}{c}{\textbf{Malay}} & \multicolumn{3}{c}{\textbf{Tamil}}\\
{Methods} & \textit{Corr} & \textit{Acc} &
  \textit{F1} &  \textit{Corr} & \textit{Acc} &
  \textit{F1}\\
\hline
RF & 0.72 & 0.71 & 0.70 & 0.43 & 0.52 & 0.59 \\
XGB & 0.75 & 0.74 & 0.73 & 0.54 & 0.65 & 0.70 \\
gpt-audio{$^\dagger$} & 0.65 & 0.71 & 0.69 & 0.43 & 0.52 & 0.51 \\
gpt-meta & \textbf{0.92} & \textbf{0.91} & \textbf{0.91} & \textbf{0.66} & \textbf{0.75} &
  \textbf{0.74} \\
\end{tabular}
}
\vspace{-3mm}
\end{center}
\end{table}

\begin{table}[tb]
\centering
\caption{Ablation study of metrics input to the tuned GPT.}
\label{tab:my-abalation study}
\scalebox{0.9}{
\begin{tabular}{lllllll}
{Excluded} & \multicolumn{3}{c}{\textbf{Malay}} & \multicolumn{3}{c}{\textbf{Tamil}}\\
{features} & \textit{Corr} & \textit{Acc} &
  \textit{F1} &  \textit{Corr} & \textit{Acc} &
  \textit{F1}\\
  \hline
wer &  0.85  & 0.81  & 0.80  & 0.61  & 0.67 &  0.67\\
cer  & 0.91  & 0.90 &  0.90  & 0.60  & 0.70  & 0.68\\
per  & 0.92  & 0.91  & 0.90  & 0.63  & 0.72 &  0.71\\
pau duration  & 0.89  & 0.87 &  0.87  & 0.64  & 0.73 &  0.73\\
tot duration  & 0.91  & 0.90 &  0.90  & 0.63 &  0.73 &  0.72\\
num pauses  & 0.91 &  0.89 &  0.89 &  0.64  & 0.74 &  0.74\\
speed &  0.91 &  0.90 &  0.90 &  0.64 &  0.74  & 0.73\\
pause ratio &  0.89 &  0.88 &  0.88 &  0.59 &  0.74  & 0.74\\
\hline
base  & 0.92 &  0.91 &  0.91 &  0.66  & 0.75 &  0.74\\
\end{tabular}
}
\vspace{-3mm}
\end{table}

Finally, comprehensive ablation experiments analyse the impact of every metric on model performance.
In Table \ref{tab:my-abalation study}, we remove each metric in turn and re-evaluate. Clearly, WER is the most important metric for Malay while CER and pause ratio are more important for Tamil. 
This may reflect the stress-timed syllabic nature of Tamil compared to the syllable-timed nature of Malay.
Interestingly, no feature appears to be redundant.

\section{Conclusion}
\label{sec:conc}

This paper has explored the automated fluency evaluation for low resource languages. We propose an automated assessment system for Malay (low resource) and Tamil (very low resource), two important languages for education in Singapore.
We adapt the powerful Whisper multilingual ASR model to extract metrics from utterances in each language and use machine learning models to provide fluency predictions from the metrics.
Our Whisper model is extensively fine-tuned with speech highly augmented by state-of-the-art methods to significantly improve baseline accuracy in each language.
For fluency evaluation, we compare two GPT systems and two machine learning methods. One GPT system is a state-of-the-art multimodal LLM directly inputting speech samples, while the other is our proposed text LLM acting as a meta-evaluator of extracted metrics.
The proposed fluency evaluation system was found to perform well in both languages, and achieves greater than 90\% accuracy in Malay.
More importantly, the proposed technique has potential to be applied to other very low resource languages in future.

\section{Acknowledgement} %
\ifinterspeechfinal
     This research is supported by the National Research Foundation, Singapore under its AI Singapore Programme (AISG Award No: AISG2-GC-2022-004). We would also like to acknowledge the contributions of the many student helpers for their assistance in this project.
\else
     This research is supported by the National Research Foundation %
     [further details are redacted for blind review]
\fi

\bibliographystyle{IEEEtran}
\bibliography{mybib}

\end{document}